\documentclass[fleqn,usenatbib]{mnras}

\usepackage[T1]{fontenc}
\usepackage{newtxtext, newtxmath}

\usepackage{tikz}
\usetikzlibrary{shapes,arrows,positioning,decorations.pathreplacing,calc}
\usepackage{rotating}
\usepackage{enumitem}
\usepackage{hyperref} 
\usepackage{xurl}

\usepackage{fancyhdr}
\title[UX Principles for Human-AI Agent Interaction]{A Framework of User Experience Principles for Human-AI Agent Interaction in the Workplace}

\author[K. Paimann et al.]{
Kathrin Paimann,$^{1}$\thanks{E-mail: kathrin.paimann@sap.com}
Elizangela Valarini,$^{1,2}$ and
Sebastian Juhl$^{1,3}$
\\
$^{1}$SAP SE, Walldorf, Germany\\
$^{2}$Hochschule Fresenius, Heidelberg, Germany\\
$^{3}$University of Missouri, Columbia, USA
}

\date{Version: \today\\
Note: This article will be published in the conference proceedings of the Mensch und Computer 2026.}

\pubyear{2026}

\begin{document}
\label{firstpage}
\pagerange{\pageref{firstpage}--\pageref{lastpage}}
\maketitle

\begin{abstract}
As AI agents become integral to business workflows, establishing guiding user experience (UX) principles is crucial for ensuring user trust and successful adoption. To address this, our study uses a multi-method approach—combining participatory design workshop, paper-and-pencil, expert review, meta-analysis, and in-depth interviews—to identify and validate a design framework of eight core UX principles for human-AI agent interaction in the workplace. Together with their underlying criteria, these principles provide actionable guardrails for designers and software engineers, creating a foundation for developing effective and human-centered AI agent interactions. This study contributes to a structured foundation for future empirical studies on agentic AI in enterprise settings.
\end{abstract}

\begin{keywords}
Human-Agent Interaction, Agentic AI, User Experience, Human-Centered AI, UX Principles
\end{keywords}

\renewcommand{\headrulewidth}{0pt}
\pagestyle{fancy}
\fancyhead{}
\fancyhead[LE]{\Large \thepage\hspace{.5cm} \textit{K. Paimann et al.}}
\fancyhead[RO]{\Large \textit{UX Principles for Human-AI Agent Interaction} \hspace{.5cm} \thepage}
\fancyfoot{}

\thispagestyle{empty}

\section{Introduction}
AI agents are emerging as a fundamental element of the modern business landscape. They are autonomous software systems that define objectives, execute multi-step plans, make decisions, interpret their operational environment, and engage with human users through natural language interaction \citep{bandi2025}. Agentic systems are occupying organizational roles previously held exclusively by humans \citep{calvanese2026, borghoff2025}. This shift fundamentally alters the conditions under which business users engage with software and the experience they expect from them. Traditional user experience (UX) practices, developed for non-agentic systems producing predictable, human-controlled outcomes, prove insufficient to account for the complexity introduced by autonomous and adaptive AI agents.

A growing body of literature addresses UX design for human interaction with AI systems, identifying criteria such as controllability, explainability, ethics, privacy, transparency, and collaboration as core design requirements \citep{xu2026, li2026, xu2024, diederich2022}. However, a systematic, empirically validated framework of UX principles with measurable criteria, specifically applicable to human-AI agent interaction in business contexts, has yet to be developed.

The present study addresses this gap by inductively deriving UX principles and measurable criteria that contribute to a positive user experience when interacting with AI agents. While the findings presented here are part of a broader research program investigating human-AI agent interaction and UX patterns in the AI era \citep{paimann2026}, this article focuses specifically on the exploratory phase of that research agenda and the derivation of the framework. Drawing on a multi-method approach which combines a secondary meta-analysis, participatory design workshops, an expert review, a paper-and-pencil survey, and semi-structured in-depth interviews, the study develops a framework that offers an empirical foundation for data-driven UX and engineering decisions in the design of agentic systems.

\section{Deriving a Framework for UX Principles in Human-AI Agent Interaction}
The evolution of Human-Computer Interaction (HCI) has progressively expanded its scope beyond traditional interface design. Early HCI scholarship concentrated on non-intelligent systems, prioritizing usability metrics, interface standardization, and task completion efficiency. Contemporary UX research recognizes this evolution as a sequence of distinct paradigms: UX 1.0 centered on personal computing and web interfaces, UX 2.0 addressing mobile and multi-device ecosystems, and the nascent UX 3.0 framework addressing AI-augmented, ecosystem-spanning experiences \citep{xu2024}. This latest paradigm introduces fundamentally new design considerations: AI systems operate with autonomous decision logic, exhibit adaptive behavior, and often employ opaque reasoning mechanisms. These characteristics engender novel risks—particularly around user trust, system controllability, algorithmic fairness, and operational safety—that extend well beyond the scope of conventional UX methodologies \citep{xu2025, shneiderman2020}.

To address these challenges, scholars have proposed foundational principles for Human-Centered AI (HCAI) design. Central to this thinking is the proposition that AI systems must function as capability augmenters rather than replacements, anchored in principles of reliability, safety, and trustworthiness, with explicit safeguards for human agency, system interpretability, and design accountability \citep{shneiderman2020}. The UX 3.0 framework extends this logic by categorizing emergent interaction modalities—ecosystem-based, innovation-driven, AI-mediated, and human-AI collaborative—and identifying cross-cutting design priorities including explainability, user control, and ethical coherence \citep{xu2024}. The HCAI Methodological Framework (HCAI-MF) operationalizes these abstractions through a hierarchical mapping from philosophical values to implementable requirements, a structured taxonomy of methods, an integrated design workflow, a model for cross-disciplinary teamwork, and tiered design governance \citep{xu2025}. While such frameworks establish valuable conceptual scaffolding, they remain largely prescriptive; they generate limited empirically grounded design specifications suited to the practical constraints of enterprise-deployed autonomous agents.

To narrow this gap, \citet{amershi2019} synthesize numerous published design recommendations into 18 actionable guidelines distributed across four interaction phases: initial encounter, ongoing interaction, error recovery, and sustained engagement. Testing across 20 commercial AI products confirms the framework's applicability and its emphasis on critical UX concerns specific to AI, mainly uncertainty communication and behavioral predictability. However, this validation focuses on consumer-oriented products. The applicability of the framework to organizational contexts remain underexplored.

Additional scholarly work in Explainable AI (xAI) converges on intelligibility as a mandatory design attribute. Yet, empirical findings remain limited since inadequately calibrated explanations frequently trigger overconfidence in agent outputs or fail to enhance decision accuracy, suggesting that design principles must account for organizational and interpersonal contexts of use, not merely interface presentation \citep{paimann2026, xu2025, shneiderman2020}.

Complementing this work, \citet{punzi2026} map the landscape of collaborative human-AI decision architectures, distinguishing three paradigm families: human oversight models, selective abstention strategies, and co-adaptive learning arrangements. These paradigms motivate designs for human-in-the-loop (HITL) interactions such as graduated escalation to expert review and reciprocal, exploratory interaction modalities. Yet the empirical foundation for these paradigms remains restricted as most evidence comes from controlled laboratory settings, and robust, longitudinal data on how these designs influence productivity, accountability, and UX within real-world business environments is scarce.

Adding to these interaction-focused perspectives, governance-oriented research by \citet{newman2023} and the \citet{nist2023} (NIST) establish structured frameworks for measuring and assuring trustworthy AI, aggregating properties such as dependability, resilience, distributional equity, privacy preservation, safety, and stakeholder accountability into architectures for design guidance. However, this literature typically targets institutional policy and risk infrastructure rather than user-facing interaction design, and empirical investigation of how trust-enhancing interventions such as confidence thresholds or decision audit logs influence user behavior and judgment within organizational settings is limited \citep{xu2025, newman2023}. Both HCAI-MF and UX 3.0 advocate for multi-dimensional measurement linking user experience metrics, business performance indicators, and organization-level outcomes. At the same time, validated measurement instruments and standardized evaluation protocols are still at a very early stage \citep{xu2025, xu2024}.

The reviewed literature reveals three key gaps that the broader research agenda to which this study belongs addresses empirically \citep[see][]{paimann2026}. First, existing frameworks are predominantly theoretical, and their proposed design criteria lack systematic empirical validation \citep{xu2025, xu2024}. Second, available guidelines are largely derived from consumer-facing AI applications and do not adequately reflect the specific demands of enterprise contexts \citep{punzi2026, amershi2019}. Third, the measurement of human-AI interaction quality remains under-specified, with no standardized metrics or assessment methodologies established to date \citep{punzi2026, xu2025}.

Through a systematic, empirically grounded approach, we derive and validate UX principles specific to human-AI agent interactions in organizational settings. By integrating these principles with existing frameworks into a coherent taxonomy and empirically establishing measurable design criteria, we propose an actionable, evidence-based foundation for the design and deployment of AI agents within business environments.

\section{Research Design}
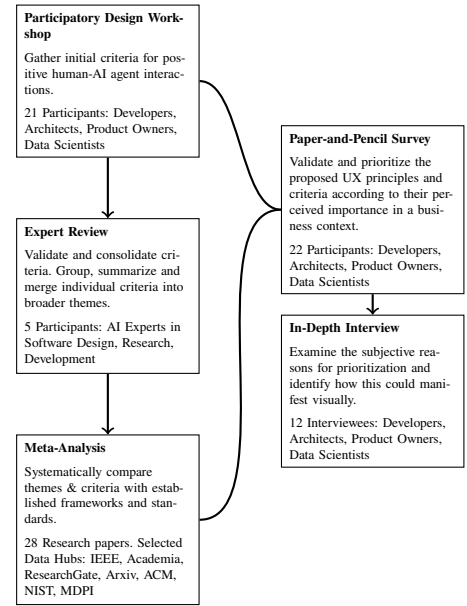
\begin{figure}
\centering
\begin{tikzpicture}[
    node distance=0.8cm,
    box/.style={
        rectangle,
        draw=black,
        text width=2.2cm,
        align=left,
        font=\tiny,
        inner sep=3pt
    },
    arrow/.style={
        ->,
        thick,
        line width=0.8pt
    }
]

\node[box] (ws) {
    \textbf{Participatory Design Workshop}\\[0.1cm]
    Gather initial criteria for positive human-AI agent interactions.\\[0.1cm]
    21 Participants: Developers, Architects, Product Owners, Data Scientists
};

\node[box, below=of ws] (er) {
    \textbf{Expert Review}\\[0.1cm]
    Validate and consolidate criteria. Group, summarize and merge individual criteria into broader themes.\\[0.1cm]
    5 Participants: AI Experts in Software Design, Research, Development
};

\node[box, below=of er] (ma) {
    \textbf{Meta-Analysis}\\[0.1cm]
    Systematically compare themes \& criteria with established frameworks and standards.\\[0.1cm]
    28 Research papers. Selected Data Hubs: IEEE, Academia, ResearchGate, Arxiv, ACM, NIST, MDPI
};

\draw[arrow] (ws.south) -- (er.north);
\draw[arrow] (er.south) -- (ma.north);

\node (vcenter) at ($(ws.center)!0.5!(ma.center)$) {};

\node[box, anchor=center] (pps) at ($(vcenter) + (3.5cm, 1.2cm)$) {
    \textbf{Paper-and-Pencil Survey}\\[0.1cm]
    Validate and prioritize the proposed UX principles and criteria according to their perceived importance in a business context.\\[0.1cm]
    22 Participants: Developers, Architects, Product Owners, Data Scientists
};

\node[box, anchor=center] (idi) at ($(vcenter) + (3.5cm, -1.2cm)$) {
    \textbf{In-Depth Interview}\\[0.1cm]
    Examine the subjective reasons for prioritization and identify how this could manifest visually.\\[0.1cm]
    12 Interviewees: Developers, Architects, Product Owners, Data Scientists
};

\draw[arrow] (pps.south) -- (idi.north);

\draw[thick] (ws.east) to[out=0, in=180, looseness=0.7] (pps.west);

\draw[thick] (pps.west) to[out=180, in=0, looseness=0.7] (ma.east);

\end{tikzpicture}
\caption{Framework development process for deriving the UX principles, including methods and samples.}\label{fig:overview}
\end{figure}

We apply a multi-method approach to empirically derive core UX principles for positive human-AI agent interactions in a business context \citep[see][]{paimann2026}. The objective is to develop a framework—built on empirically validated principles and criteria for human-AI agent interaction—that provides an advanced understanding of this complex, multidimensional concept as well as practical guidance for developing human-centered AI agent interactions. Figure \ref{fig:overview} provides a visual overview on how the UX principles are identified and validated, including information on the methodologies and sample.

\subsection{Participatory Design Workshop and Expert Review}
The data collection started with a participatory design workshop to generate an initial draft of UX principles and criteria for a positive Human-AI agent interaction experience. The 60 minute in-person session brought together participants with prior AI experience and AI agent familiarity, organized into role-based groups, each tasked with identifying criteria for a positive interaction with AI agents. Participants' knowledge and articulated expectations, requirements, and perceived risks served as the primary epistemic source for exploring UX criteria for agent interaction and building the framework \citep{arcia2024}. The findings were clustered, reviewed, and systematically structured by five domain experts in AI to identify recurring topics.

\subsection{Meta-Analysis of Secondary Literature}
To develop the framework, we conducted a qualitative meta-analysis \citep[see e.g.,][]{shen2024, sandelowski2007} and aggregated the synthetized findings from the participatory design workshop. For the meta-analysis, we constructed and analyzed a corpus of selected peer-reviewed publications and one governance framework. Keywords such as HCAI, HCI, HAI, agents, frameworks, principles, and business context serve as selection criteria \citep[see][]{paimann2026}. The corpus encompasses the following publications:

\begin{itemize}[leftmargin=0.5cm]
\item Bradshaw at al. (2011) Human-Agent Interaction.
\item Diederich at al. (2022) On the Design of and Interaction with Conversational Agents: An Organizing and Assessing Review of Human-Computer Interaction Research.
\item Nastoska at al. (2025) Evaluating Trustworthiness in AI Risks, Metrics, and Applications Across Industries.
\item NIST (2023) AI Risk Management Framework 1.0. 
\item Shneiderman, B. (2020) Human-Centered Artificial Intelligence: Reliable, Safe \& Trustworthy.
\item Xu, W. (2024) A User Experience 3.0 Paradigm Framework.
\item Xu at al. (2025) An HCAI Methodological Framework \mbox{(HCAI-MF):} Putting It Into Action to Enable Human-Centered AI.
\end{itemize}

The synthesized findings from the participatory design workshop were used to derive the initial higher-order categories. The corpus was subsequently analyzed qualitatively, and the results were integrated with the initial categories derived from the workshop. The meta-analysis consisted, first, of identifying common principles and criteria across all selected papers, and second, of identifying exceptions—that is, criteria appearing in only a single source. In a subsequent step, we excluded principles unrelated to UX, such as technical aspects or requirements beyond the scope of UX design. This analysis yielded eight high-level UX principles, each associated with 3-6 underlying criteria. Methodological differences between the analyzed literature were not considered in the analysis and synthesis of the principles.

\subsection{Paper-and-Pencil Survey}
Using the eight core UX principles and 3-6 criteria for positive human-AI agent interaction identified by the two methodological approaches discussed above as a basis, we employ a paper-and-pencil survey to validate and prioritize the principles and criteria according to end-user and expert requirements. The survey was conducted during a workshop, with 45 minutes allocated for completion and 15 minutes for a subsequent group activity. The paper format is particularly suitable in this in-person workshop setting, as it yields higher response rates and less extreme responses than computer-based surveys \citep{colasante2019, lewis2009}. Participants had the option to provide additional open-ended feedback before discussing their rankings in role-specific groups.

\subsection{In-Depth Interviews}
To gain further insights into the prioritization of the UX principles, we conduct twelve qualitative, semi-structured in-depth interviews virtually. To recruit interviewees, we used an online recruiting platform, while workshop participants were recruited separately. All interviewees had professional experience with AI tools and agent-based technologies. They were experts in software development, data science, and products across diverse industries. The semi-structured interviews aimed to provide a deeper understanding of survey participants' prioritization by explaining the meaning and reasoning underlying their choices and by exploring specific requirements and identifying recurring patterns \citep{brinkmann2020, kvale1996}. The interview guide comprised core open-ended questions focusing on participants' top three ranked UX principles, progressing from broad context-setting to more specific probes, with the flexibility to follow up on emergent themes. Our analysis follows a deductive-inductive thematic approach which combines pre-defined categories with newly emerged themes \citep{braun2006}.

\section{A Framework for Human-AI Agent Interactions Consisting of Eight UX Principles}
The analysis and triangulation of empirical data collected across the multi-method research phases yielded a validated framework of eight agent UX principles, ranked by their perceived importance in a business context . Each principle is defined by 3-6 underlying criteria, which are formulated as specific, measurable, and actionable rules for successful design implementation. This section introduces the derived UX principles in greater detail and presents the findings from the validation study. Table \ref{tab:principles} presents the principles.\\

\begin{table}
\tiny
\centering
\rotatebox{00}{
\begin{tabular}{lp{.85\linewidth}}
\hline
1 & \textit{A human is always in control of an agent} \\
\hline
& \vspace{-\baselineskip}\begin{itemize}[leftmargin=*, align=parleft, topsep=0pt, partopsep=0pt, itemsep=0pt]
\item A human maintains ultimate authority, with the ability to intervene, override, or disengage the agent.
\item A human is always responsible for the decisions made in relation to an agent's actions.
\item A human always has to confirm before an agent takes a critical (business) decision.
\item A human is always aware of the agent's intended steps or actions before it executes them.
\item An agent works autonomously only on low-impact, time-consuming and/or manually intensive tasks.
\end{itemize}\\
\hline
2 & \textit{An agent operates transparently and explains its output} \\
\hline
& \vspace{-\baselineskip}\begin{itemize}[leftmargin=*, align=parleft, topsep=0pt, partopsep=0pt, itemsep=0pt]
\item An agent provides clear reasoning and describes ``how'' and ``why'' it came to a result.
\item An agent's actions and outputs can be easily verified.
\item An agent's output is understandable for the user’s specific skill level and context.
\item An agent provides sufficient information on how it works, including the rules or logic it follows.
\item A human can easily determine an agent's current status and progress.
\item An agent provides completion reports to confirm accomplishment.
\end{itemize}\\
\hline
3 & \textit{An agent is reliable, safe and robust} \\
\hline
& \vspace{-\baselineskip}\begin{itemize}[leftmargin=*, align=parleft, topsep=0pt, partopsep=0pt, itemsep=0pt]
\item An agent's output is accurate and avoids hallucinations or misleading information.
\item An agent always uses up-to-date resources and data.
\item An agent provides clear risk indications (e.g., knowledge limits or uncertainty).
\item An agent clearly shows which human input is needed (e.g., decision).
\item When an agent fails to understand an input, it applies ``repair strategies'' that are actionable and explanatory.
\item An agent implements audit trails to reveal its process (e.g., activity log).
\end{itemize}\\
\hline
4 & \textit{An agent is always context-aware} \\
\hline
& \vspace{-\baselineskip}\begin{itemize}[leftmargin=*, align=parleft, topsep=0pt, partopsep=0pt, itemsep=0pt]
\item An agent is aware of the user's role and permissions when providing relevant assistance.
\item An agent provides guidance and support for multi-stage processes or tasks, maintaining context from one step to another.
\item An agent understands questions in the context of recently uploaded data, enabling it to provide accurate answers and relevant suggestions.
\item An agent adapts over time to the personal working style of a user.
\item An agent is aware of the user's team structure and context.
\end{itemize}\\
\hline
5 & \textit{An agent acts as a collaborative partner} \\
\hline
& \vspace{-\baselineskip}\begin{itemize}[leftmargin=*, align=parleft, topsep=0pt, partopsep=0pt, itemsep=0pt]
\item A human primarily interacts with a single ``orchestrating'' agent that coordinates a team of specialized agents and reroutes queries.
\item A human can add and remove specialized agents from an interaction, task or process.
\item A human is always aware of all specialized agents involved in an interaction even if s/he does not directly interact with them.
\item An ``orchestrating'' agent acts like a ``teammate'' that performs tasks while asking for approval or clarification at key moments.
\end{itemize}\\
\hline
6 & \textit{An agent adheres to data privacy \& data governance} \\
\hline
& \vspace{-\baselineskip}\begin{itemize}[leftmargin=*, align=parleft, topsep=0pt, partopsep=0pt, itemsep=0pt]
\item An agent enforces strict data privacy protection.
\item An agent enforces role-based access controls to ensure the output is appropriate for the user's permissions.
\item Humans provide consent over how their data is collected and used.
\item An agent operates predictably by adhering to data governance protocols.
\item An agent follows robust security standards to protect the system and the data it accesses.
\end{itemize}\\
\hline
7 & \textit{An agent is integrated into the ecosystem} \\
\hline
& \vspace{-\baselineskip}\begin{itemize}[leftmargin=*, align=parleft, topsep=0pt, partopsep=0pt, itemsep=0pt]
\item An agent has access to the full ecosystem of enterprise applications, including sap and third-party systems.
\item An agent functions as a collaborative process orchestrator to manage complex workflows from start to finish.
\item An agentic system can access multiple apps seamlessly and abstract background complexity into a simple user interface.
\end{itemize}\\
\hline
8 & \textit{An agent is responsive and intuitive to use} \\
\hline
& \vspace{-\baselineskip}\begin{itemize}[leftmargin=*, align=parleft, topsep=0pt, partopsep=0pt, itemsep=0pt]
\item An agent can be used seamlessly across devices.
\item The primary interaction between the human and agent is a conversational UI that adheres to UX best practices (i.e., accessibility, consistency, intuitiveness etc.).
\item An agent communicates in natural language.
\item A human can easily determine an agent’s current status and progress.
\end{itemize}
\end{tabular}
}
\caption{UX principles and the underlying criteria for human-AI agent interaction design.}\label{tab:principles}
\end{table}

\noindent
{\bf A Human is Always in Control an of Agent.} The principle of human control is ranked as most important, with 65\% of paper-and-pencil survey participants choosing it as one of their top three most important UX principles and 40\% ranking it as their number one. Most participants (95\%) indicate that a human must always confirm before an AI agent takes a critical business decision. During the interview, participants named examples of critical business decisions participants such as decisions that require budget above a defined limit, decisions with an impact on business results, decisions that have a long-term impact such as supplier selection, and decisions that have an impact on employees. Furthermore, 86\% of survey the participants require that a human maintains ultimate authority, with the ability to intervene, override, or disengage the AI agent. Control is perceived as being the initiator and guide, without needing to micromanage every single step. In total, 40\% of participants require that a human always remains accountable for an agent's decisions and actions. For the design of agentic interactions, this implies that a user needs the option to stop or pause an agent any time. Human input is always required for critical business decisions prior to execution. Moreover, users need to be able to understand the implications of their decision and be aware of their accountability.\\

\noindent
{\bf An Agent is Reliable, Safe and Robust.} In total, 60\% of the paper-and-pencil survey participants chose this UX principle as one of their top three most important. The highest rated UX criteria includes the requirement that an agent's output is accurate and avoids hallucinations or misleading information (65\%). Another 57\% of participants require an agent to always use up-to-date resources and data, and provide clear risk indications, for example, regarding uncertainty. This principle was seen for most participants as a non-negotiable baseline in agent interactions. For the design of agentic interactions, this means that the use of confidence levels, indicators for knowledge limits, and information on when data was last updated are helpful.\\

\noindent
{\bf An Agent Adheres to Data Privacy \& Data Governance.} The ranking shows that 45\% of participants select this UX principle as one of their top three. A majority of participants (85\%) require an agent to enforce strict role-based access controls to ensure the output is appropriate for the user's permissions. Additionally, 76\% of participants require that an agent enforces strict data privacy protection and adheres to robust security standards to protect both the system and the data it accesses. The risk of mishandling confidential company or client data was seen as a disastrous failure with potentially irreversible legal and financial consequences. For design purposes, these findings suggest that sensitivity labels are beneficial when sensitive data is accessed or about to be released. The system should also clearly inform the user when role-based permissions are lacking and potentially provide an option to request access.\\

\noindent
{\bf An Agent is Always Context-Aware.} Similarly to the previous UX principle, this one is selected by 45\% of participant as one of their top three. A majority of participants (95\%) require an agent to be aware of the user's role and permissions when providing relevant assistance. Another 72\% need an agent to provide guidance and support for multi-step processes or tasks, maintaining context from one step to another and offering accurate answers and relevant suggestions based on recently uploaded data. This principle is seen as the key differentiator from a generic chatbot. Interviewees indicate that in a business context, this principle is what makes an agent's output usable and relevant since outputs without context are often too unspecific to be actionable. For the design of agentic interactions, a clear indication of lacking context can be helpful, as well as indicators showing whether content has been added or reviewed by a human or an AI agent.\\

\noindent
{\bf An Agent Operates Transparently and Explains its Output.} In total, 40\% of participants choose this principle as one of their top three. The most important criteria are that an agent's actions and output can be easily verified (75\%) and that a clear reasoning and description of ``how'' and ``why'' an agent came to a result is necessary (60\%). Additionally, 55\% of participants require an agent to provide sufficient information on how it works, including the rules and logic it follows. Transparency is the foundation for building trust and enabling human control. Users need to understand not just what the agent did, but why it did it, especially when the outputs are complex or have far-reaching consequences. At the same time, we find that the need for detailed explanations is contextual. While it is less important for simple, transactional tasks, it is critical for content creation or complex problem-solving. For novice users, transparency seems to be more relevant to build trust, whereas experienced users prioritize speed over transparency. For design purposes, it is beneficial to allow for the verification of information sources and data consumed by the agent.\\

\noindent
{\bf An Agent is Integrated into the Ecosystem.} 40\% of participants select this principle as one of their top three. As key requirement, 40\% of participants want an agentic system that can access multiple apps seamlessly and abstract background complexity into a simple user interface. Another 31\% of participants want an agent to function as a collaborative process orchestrator to manage complex workflows end-to-end. Integration is seen as a powerful enabler of efficiency and context. As such, this principle is viewed as deeply connected to context-awareness, as it provides the agent with the necessary information to understand the environment it operates in. A standalone agent that requires manual data uploads is considered as far less valuable than an agent that can seamlessly pull information from different tools and knowledge bases.\\

\noindent
{\bf An Agent Acts as a Collaborative Partner.} With 15\% of participants choosing this principle as one of their top three, it is selected less frequently compared to the previous principles. The vast majority of participants (95\%) require that a human can add and remove specialized agents from a workflow, task, or process. Another 85\% of participants want an agent that acts like a teammate, performing tasks while asking for approval or clarification at key moments. Interviewees mention they want an agent that functions as a true partner that supports their workflow, anticipates needs, and helps them achieve their goals more efficiently, but without overstepping its role. The agent should support the human and the process, not replace human judgment. For the design, this implies that agents need to be clearly identifiable as such.\\

\noindent
{\bf An Agent is Responsive and Intuitive to Use.} This principle is the least frequently selected. Only 35\% of participants currently rank a conversational UI that adheres to UX best practices such as accessibility, consistency, and intuitiveness as the primary interaction format with agents amongst the top 3 UX principles. Through the interviews we find that ease of use is a key driver of adoption. If an agent is slow, clunky, or difficult to interact with, users will abandon it, regardless of the quality of its output.

\section{Conclusion}
This study contributes to the HCI and HCAI literature by providing an empirically grounded, enterprise-focused foundation for designing human-AI agent interactions. Through a rigorous multi-method approach, we identified and systematized eight core UX principles with actionable underlying criteria and derived a design framework to narrow the critical gap between rapidly advancing agentic AI capabilities and their adaptation in organizational contexts.

Our framework addresses a fundamental challenge facing enterprises today: how to integrate AI agents into workflows while maintaining meaningful human oversight, fostering user trust, and adhering to organizational governance requirements. Our results show that business users prioritize principles centered on human control, reliability, context-awareness, and safety. These are not merely theoretical ideals but practical necessities that directly shape user acceptance and effective collaboration with AI agents.

Furthermore, the underlying criteria derived here provide concrete, implementable guidance for UX designers and software engineers. Rather than merely offering abstract design principles, our framework translates user expectations into specific design decisions that preserve human agency while enabling AI agents to function effectively. Thereby, we provide a foundation for consistent, human-centered AI agent development.
Our empirical findings highlight that there is no one-size-fits-all approach to human-AI agent interaction. The principles and criteria identified reflect the complex, contextual nature of enterprise work, where factors such as job role, task criticality, decision impact, and organizational constraints shape user needs. This underscores the importance of context-aware designs in agentic systems.

At the same time, some limitations of the present research need to be considered when applying the framework. First, the eight principles identified are not mutually exclusive. Conceptual overlaps exist between some principles which reflects the reality of how these concerns manifest in practice and should be acknowledged by practitioners implementing the framework. Second, the participatory design workshop involved only a relatively small number of self-selected participants. This sample, while valuable for generating rich qualitative insights, may not represent the full diversity of enterprise users or organizational contexts. Workshop participants may have held particular perspectives on AI agent interaction that differ from the broader population of business users, potentially limiting the external validity of the findings. Third, this design enforced strict prioritization of the principles by requiring participants to rank them without the option of assigning equal importance to more than one principle, even when multiple principles were perceived as equally important. Additionally, the framework’s validation was conducted in controlled research settings; real-world deployment contexts may present unforeseen complexities.

Future research should explore how these principles and criteria manifest differently across industry sectors, organizational structures, and task domains. We particularly recommend empirical studies investigating heterogeneous effects across user subgroups and examining alternative design implementations of each criterion. Furthermore, longitudinal studies examining how user trust and adoption of AI agents evolve over time would significantly enrich our understanding. Finally, extending this framework to domain-specific contexts (e.g., healthcare, finance, creative industries) to probe the framework’s scope conditions remains a critical avenue for validation and refinement.

By establishing this structured foundation, this research enables both immediate practical application and provides a common foundation for a broader research agenda on designing trustworthy, effective human-AI agent interactions in enterprise environments.


\section*{Acknowledgement}
We would like to thank two anonymous reviewers for their input. Special thanks to Dr. Giannis Misiakos, Linda G\"otz, Lara Valenti, Jennifer Shore, Ana Tarasova, Carsten Schmitt, Diego Ferrin, Martin Schrepp, Manvi Verma and Marina Kellermann for their contribution to this research project.



\bibliographystyle{mnras}
\bibliography{bibliography}



\label{lastpage}
\end{document}